# Exploring the Universe via the Wide, Deep Near-infrared Imaging ESO Public Survey SHARKS

Helmut Dannerbauer[1,2]
Aurelio Carnero[1,2]
Nicholas Cross[3]
Carlos M. Gutierrez[1,2]

[1] Instituto de Astrofísica de Canarias, Tenerife, Spain
[2] Department of Astrophysics, University of La Laguna, Tenerife, Spain
[3] Institute for Astronomy, University of Edinburgh, Royal Observatory Edinburgh, UK

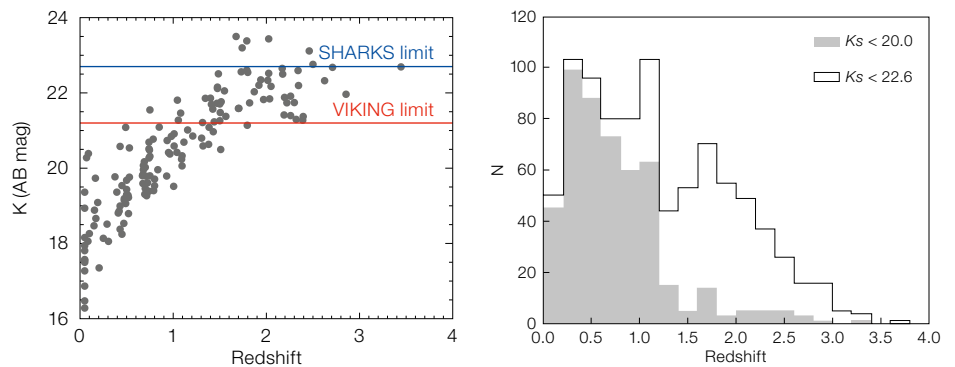

Figure 1. Left: $Ks$-band magnitude as a function of redshift for Herschel sources (from HerMES; Oliver et al., 2010) with Very Large Array counterparts in the Subaru Deep Field and the COSMOS survey. The VIKING survey is only able to detect H-ATLAS galaxies at $z < 1$. The much deeper SHARKS observations will enable detection of 90% of all star-forming, obscured, H-ATLAS sources, and also the most extreme populations at $z > 3$. Right: Photometric redshift distribution of counterparts of simulated SKA sources in the VIDEO survey, from McAlpine et al. (2012). This Figure shows that shallow $Ks$-band surveys only select the $z < 1$ radio population, whereas deeper surveys such as SHARKS are absolutely needed to select the $1 < z < 3$ population and the most extreme sources at $z > 3$.

The ESO Public Survey Southern H-ATLAS Regions $Ks$-band Survey (SHARKS) comprises 300 square degrees of deep imaging at 2.2 microns (the $Ks$ band) with the VISTA InfraRed CAMera (VIRCAM) at the 4-metre Visible and Infrared Survey Telescope for Astronomy (VISTA). The first data release of the survey, comprising 5% of the data, was published via the ESO database on 31 January 2022. We describe the strategy and status of the first data release and present the data products. We discuss briefly different scientific areas being explored with the SHARKS data and conclude with an outline of planned data releases.

## Why another near-infrared survey?

The near-infrared (near-IR) wavelength range, a type of light imperceptible to the human eye, allows us to explore regions of the Universe that are either obscured by cosmic dust, too cold to be studied with telescopes observing in the visible, or at sufficiently high redshift that rest-frame optical/ultraviolet light is visible in the near-IR. Thus, this spectral range is key to understanding galaxies at both low and high redshifts. New wide-area surveys in the far-IR and radio promise to revolutionise this field of research, but currently we lack equivalent deep, wide-area, 2.2-micron ($Ks$-band) imaging (the most efficient near-IR passband in which to observe distant galaxies) to link the radio/far-IR sources to the optical and near-IR.

The Herschel ATLAS (H-ATLAS; Eales et al., 2010) is the largest Herschel Space Observatory open-time key project, having detected approximately half a million sources up to $z = 5$ with $S_{250micron} > 28$ mJy over 600 square degrees of the sky. The discovery of distant, dust-obscured, star-forming galaxies in particular triggered a large number of pointed multi-wavelength campaigns. Clearly, at the depth of the VISTA Kilo-degree Infrared Galaxy (VIKING) Public Survey[a] (the deepest near-IR survey available over the South Galactic Plane (SGP) and Galaxy And Mass Assembly (GAMA) survey areas), only H-ATLAS sources at $z < 1$ have near-IR counterparts (Figure 1). However, more than 90% of the H-ATLAS sources at $1 < z < 3$ would be detected down to $Ks = 22.7$ mag (AB). Figure 1 also shows the redshift distribution of simulated Square Kilometre Array (SKA) radio sources in two different $Ks$-band magnitude ranges. It can be seen that shallow $Ks$-band observations/surveys such as VIKING only detect the $z < 1$ radio population, whereas surveys as deep as the VISTA Deep Extragalactic Observations (VIDEO; Jarvis, Häußler & McAlpine, 2013) will also detect the obscured sources at $1 < z < 3$ and some of the most extreme systems at $z > 3$. Therefore, the lack of wide and deep near-IR imaging on H-ATLAS fields motivated our team to propose a new public survey in 2015.

## SHARKS survey and observations

The ESO Public Survey Southern H-ATLAS Regions $Ks$-band Survey (SHARKS; PI: H. Dannerbauer; 198.A-2006) was approved as a second-generation survey with the 4-metre near-IR-optimised Visible and Infrared Survey Telescope for Astronomy (VISTA) and its near-IR, wide-field imager the VISTA InfraRed CAMera (VIRCAM) at ESO's Paranal Observatory. In total, the survey was granted 1200 hours planned initially over four years. The overall survey covers 300 square degrees down to a 5-sigma limit of $Ks = 22.7$ mag (AB-system, 2-arcsecond aperture) and will detect a total of approximately 20 million sources. Already more than 90% of the data has been taken since autumn 2016 and it is expected that the observations will be completed by the end of 2022. The SHARKS team is an international collaboration of more than 50 researchers, whose expertise covers all aspects related to the full and successful exploitation of the SHARKS dataset.

In all, we will have conducted roughly 200 pointings (mosaics) in three fields, GAMA12, GAMA15 and SGP (see Figures 2 and 3). Each pointing is visited seven times (seven observing blocks of ~ 55 minutes' duration each). Within each observing block 36 images are taken, each with 60 seconds integration time, with the rest of the observing block dedicated to overheads. The survey strategy aims to complete each mosaic before moving to another, therefore the epoch difference within a mosaic is in general around a month. The required sky conditions are: seeing < 1.2 arcseconds in the SGP field and < 1.0 arcseconds in the GAMA fields, airmass < 1.7 and clear sky.

### Data reduction

The first SHARKS data release (DR1) has been produced by a collaboration between the Instituto de Astrofísica de Canarias (IAC) and the Wide-Field Astronomy Unit (WFAU) at the Royal Observatory Edinburgh. Images are processed and calibrated at the WFAU. In addition, we have implemented an improved sky subtraction method. We use images reduced at the Cambridge Astronomy Survey Unit (CASU) and corrected for reset, dark, linearity, flat-field, and sky-background stripe correction. The VISTA Data Flow System pipeline processing and science archive are described by Irwin et al. (2004), Hambly et al. (2008) and Cross et al. (2012). Details of the data processing and calibration are provided in a separate document describing DR1[1]. To summarise, the pipeline starts by retrieving the CASU processed images and ends by creating co-added mosaic images sampled to a pixel size of ~ 0.34 arcseconds. The images have been astrometrically and photometrically calibrated with respect to the Two Micron All Sky Survey (2MASS; Skrutskie et al., 2006) and source catalogues are obtained with the SExtractor code (v2.19.5; Bertin & Arnouts, 1996).

### First data release

On 31 January 2022 we made the DR1, amounting to 5% of the SHARKS data, available to the scientific community via the ESO Science Archive Facility. The release contains both calibrated images and source catalogues, so the data can be immediately exploited by researchers worldwide. The images from the DR1 are distributed over the three SHARKS fields, where six of them are isolated and four of them cover a contiguous region in the SGP-East (SGP-E) field (see Figure 3). The total area of each mosaic is 2.03 square degrees. In the four contiguous pointings in the SGP-E field, there is an overlap of 1.03 square degrees (13%) within the mosaics, completing 7.06 square degrees (see Figures 3 and 4). In total, the unique area of DR1 is 19.24 square degrees (20.27 square degrees considering the overlap). The DR1 consists of 10 co-added images and individual $Ks$-band source catalogues. Photometry is given for 13 standard apertures, from 1 to 24 arcseconds, in the AB system. For extended sources, Kron (Kron, 1980) and Petrosian (Petrosian, 1976) fluxes — which account for much of the missing light — are also determined. In addition to the co-added images we provide normalised-weight images (normalised to effective gains in analogue-to-digital units) and pre-images related to each (weight) image. The SHARKS DR1 catalogue comprises more than 1.5 million sources down to a median depth of 22.7 (5-sigma, AB), and a median seeing of 1 arcsecond. We estimate that 90% are classified as galaxies and that the level of spuriousness is below 2%, concentrated around bright stars. The release is made up of ~ 18 GB of co-added images (including weights) and 700 MB of the catalogues, provided individually for each of the 10 mosaics. The document giving details of the released data[1] also describes data reduction and quality.

### Legacy value

The SHARKS dataset is complemented by an excellent existing multi-wavelength dataset including observations from the Spitzer Space Telescope, the Atacama Large Millimeter/submillimeter Array (ALMA), the Australian SKA Pathfinder (ASKAP), the Low Frequency Array (LOFAR), the Dark Energy Survey (DES), the Subaru Telescope and the Hubble Space Telescope (see Figure 2). These data are a mix of individual pointed observations and large-area surveys. The SHARKS fields covered will also overlap with future Legacy Survey of Space and Time (LSST; optical) and Euclid (near-IR, but not $Ks$) deep observations, representing a perfect complementary dataset (see Figure 2). To summarise, the combination of the near-IR survey SHARKS with previous, current and future surveys at different wavelengths has a unique legacy value for future astrophysical studies.

Figure 2. Coverage of the SHARKS fields by current and future surveys and missions.

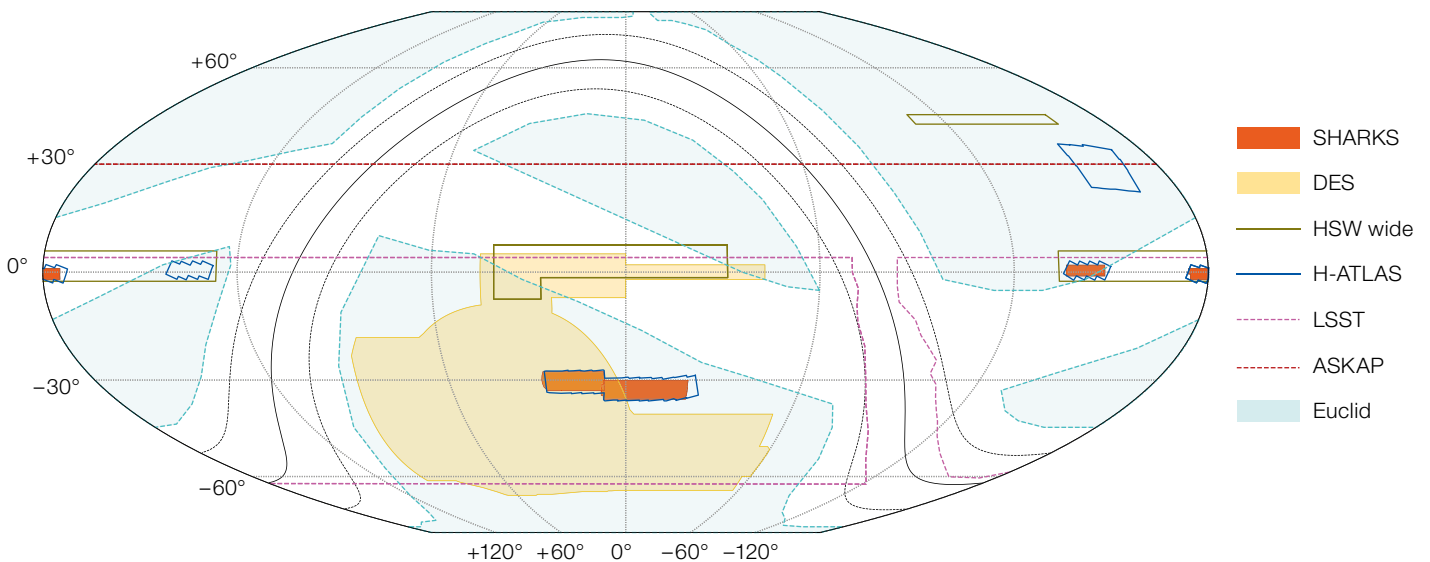

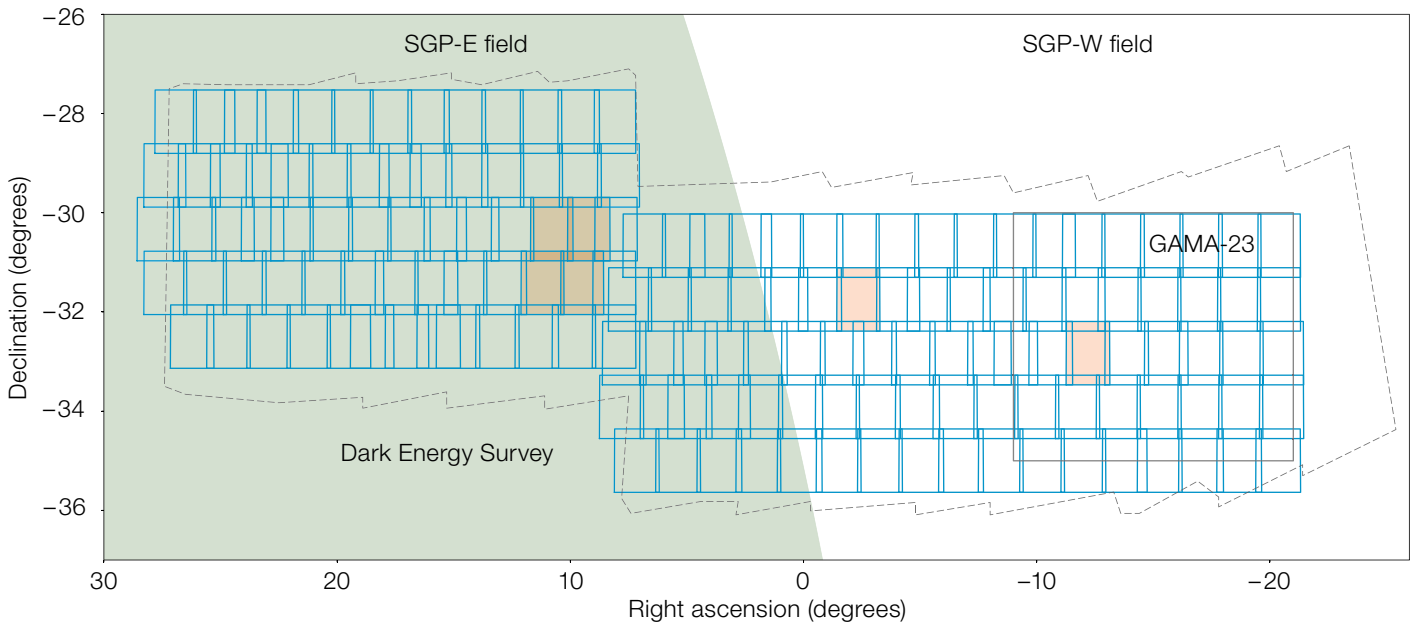

Figure 3. Spatial distribution of all observed and planned mosaics (in blue) in the SGP field. The H-ATLAS footprint is shown dashed, the GAMA-23h footprint is in black and the DES foorprint is shown filled in green. In red we show the mosaics of this field that were released in DR1.

### Science outlook

The SHARKS data can be explored in different astrophysical areas, such as Solar System objects, stars, local and high-redshift galaxies and large scale structures. Thanks to the wide angular coverage of SHARKS, cosmic variance is negligible, and thus statistical studies of the evolution and formation of galaxies will provide robust and reliable conclusions. Using the four contiguous SGP-E fields from the DR1, we have derived the $Ks$-band number counts down to $Ks = 22.7$ mag (see Figure 5)[b]. This estimate is consistent with previous work by Daddi et al. (2000), giving us confidence in our data reduction and calibration. The larger part of the SHARKS survey has coverage in the optical from the Hyper Suprime-Cam Subaru Strategic Program (HSC-SSP) and the DES. Adding the SHARKS imaging already improves photometric redshifts estimates for high-$z$ sources significantly. In future, LSST and Euclid imaging will certainly further improve these estimates. The following topics related to galaxies can be exploited and searched for with the SHARKS data: the far-IR/radio correlation, rare objects, galaxy overdensities, lensed galaxies and QSOs. As shown in Figure 1, for about 90% of Herschel-selected sources, SHARKS will provide a counterpart identification, indispensable for deriving their distances, stellar masses and star formation rates. This is also true for sources selected from LOFAR and ASKAP radio surveys and ALMA observations. Currently the only region of sky with $Ks$-band imaging sufficiently deep to provide this level of identification completeness for far-IR/radio surveys is VIDEO. But at only 12 square degrees there will be little scope for probing extreme and rare sources. With 25 times more area, to the $Ks$-band depth required for radio identification, SHARKS would be ideal for finding the highest-redshift powerful radio sources with baryonic masses of $10^{12} M_\odot$ to $z = 10$ (Rocca-Volmerange et al., 2004).

### Future plans

In the coming years at least two more ESO data releases are planned. In the second one, aimed for the beginning of 2023, we plan to release at least 50% of the data. In the third (and probably final) one, we will release all the observed fields. We also plan to include multi-wavelength information for each SHARKS $Ks$-band source. For future data releases we aim to use Gaia catalogues for astrometric calibration. In addition, the SHARKS team members at the IAC are working on a specific data reduction in order to reveal low-surface-brightness objects — such as Galactic cirrus, ultra-faint galaxies, outer parts of galaxies and intracluster light — in the near-IR. This work is based on previous work in the optical (Trujillo & Fliri, 2016) and will give us access to astronomical objects previously unexplored in this wavelength regime. Finally, we would like to note that more information about the survey, including the full list of SHARKS co-investigators, is available on our dedicated SHARKS webpage[2].

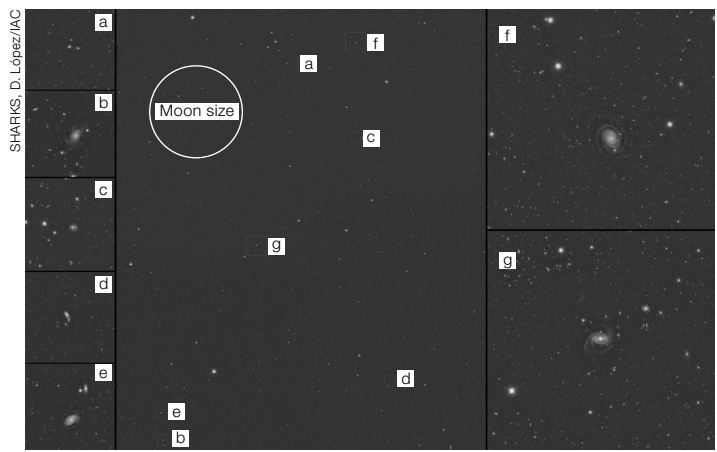

Figure 4. Near-IR image of the SHARKS survey of a 7-square-degree large field (the four contiguous SGP-E fields from DR1), with the apparent size of the Moon as a reference. Boxes present representative galaxies in detail.

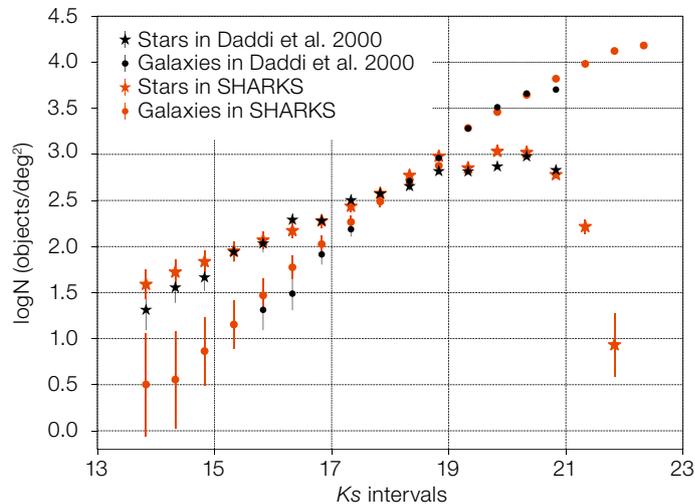

Figure 5. Comparison of $Ks$-band number counts of stars and galaxies derived for the four contiguous fields in SGP-E from SHARKS DR1 and work presented in Daddi et al. (2000). Our galaxy number counts match well with the previous study and we could even extend them to fainter magnitudes.

### Acknowledgements

The authors acknowledge Ivan Oteo, with sincere thanks for the early impetus he gave this project. Based on data products created from observations collected at ESO under programme 198.A-2006. For the creation of the data used in this work, the SHARKS team at the Instituto de Astrofísica de Canarias has received financial support from the Spanish Ministry of Science, Innovation and Universities (MICIU) under grant AYA2017-84061-P, co-financed by FEDER (European Regional Development Funds), from the Spanish Space Research Program "Participation in the NISP instrument and preparation for the science of EUCLID" (ESP2017-84272-C2-1-R), and from the Agencia Canaria de Investigación Innovación y Sociedad de la Información Gobierno de Canarias (ACIISI), the Consejería de Economía, Conocimiento y Empleo del Gobierno de Canarias and the European Regional Development Fund (ERDF) under grant PROID2020010107. We thank the Wide-Field Astronomy Unit for testing and parallelising the mosaic process and preparing the releases. The work of the Wide-Field Astronomy Unit is funded by the UK Science and Technology Facilities Council through grant ST/T002956/1.

### References

Bertin, E. & Arnouts, S. 1996, A&AS, 117, 393
Cross, N. J. G. et al. 2012, A&A, 548, A119
Daddi, E. et al. 2000, A&A, 361, 535
Eales, S. et al. 2010, PASP, 122, 499
Edge, A. et al. 2013, The Messenger, 154, 32
Hambly, N. C. et al. 2008, MNRAS, 384, 637
Irwin, M. J. et al. 2004, Proc. SPIE, 5493, 411
Jarvis, M. J., Häußler, B. & McAlpine, K. 2013, The Messenger, 154, 26
Kron, R. G. 1980, ApJS, 43, 305
McAlpine, K. et al. 2012, MNRAS, 423, 132
Oliver, S. J. et al. 2010, A&A, 518, L21
Petrosian, V. 1976, ApJL, 209, L1
Rocca-Volmerange, B. et al. 2004, A&A, 415, 931
Skrutskie, M. F. et al. 2006, AJ, 131, 1163
Trujillo, I. & Fliri, J. 2016, ApJ, 823, 123

### Links

[1] DR1 description: https://www.eso.org/rm/api/v1/public/releaseDescriptions/179
[2] SHARKS webpage: http://research.iac.es/proyecto/sharks/pages/en/home.php

### Notes

[a] The H-ATLAS fields were covered previously by the ESO Public Survey VIKING (Edge et al., 2013) at a significantly shallow level ($Ks < 21.2$ mag, 5-sigma, AB).
[b] This analysis was part of a bachelor's thesis carried out by Sergio Saavedra Esquiviel of the Universidad de La Laguna. For details see: https://riull.ull.es/xmlui/handle/915/24098

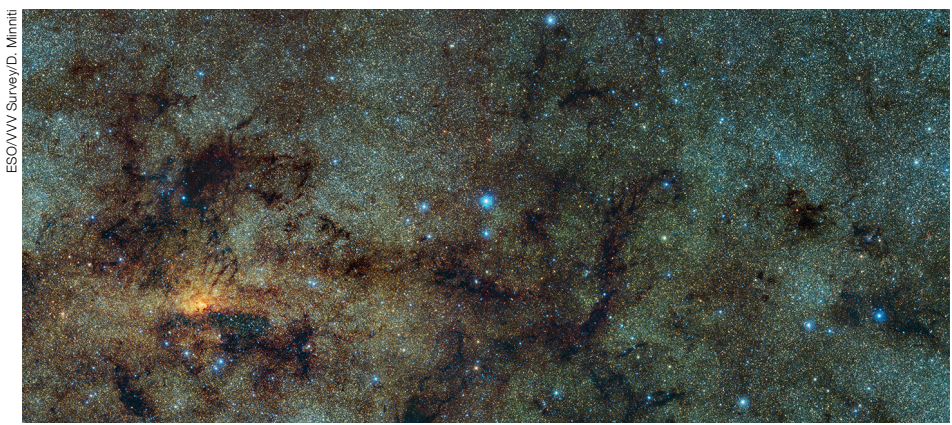

This image, captured with the VISTA infrared survey telescope, as part of the Variables in the Via Lactea (VVV) ESO public survey, shows the central part of the Milky Way. While normally hidden behind obscuring dust, the infrared capabilities of VISTA allow the stars close to the galactic centre to be studied.